\documentstyle[12pt,psfig]{ioplppt}

\begin{document}
\jl{1}
\bibliographystyle{plain}

\title{Droplets in the coexistence region of the two-dimensional Ising model}[Droplets in Ising model]
\author{M Pleimling\S\dag ~ and ~ W Selke\S}
\address{\S\ Institut f\"ur Theoretische Physik, Technische
Hochschule, D--52056 Aachen, Germany}
\address{\dag\ Institut f\"ur Theoretische Physik 1, Universit\"at Erlangen-N\"urnberg, D--91058 Erlangen, Germany}

\begin{abstract}
The two--dimensional Ising model with fixed magnetization is
studied using Monte Carlo techniques. At the coexistence
line, the macroscopic, extensive droplet of minority spins becomes
thermally unstable by breaking up into microscopic
clusters. Intriguing finite--size effects as well as singularities
of thermal and cluster properties associated with the transition
are discussed.
\end{abstract}
\maketitle

%\section{Text}
The two--dimensional Ising model has attracted much interest in
the past. It is conceptually simple, and several of its non--trivial
properties can be determined exactly \cite{mccoy}. However, despite
numerous studies, the model seems to be not yet completely
understood.

In this Letter, we shall deal with the thermal stability of a
droplet of minority spins in the two--dimensional Ising
model with {\it fixed magnetization}, monitoring the transition
from a compact single, macroscopic and extensive droplet at low
temperatures to an ensemble of small clusters at
high temperatures. Although related cluster
equilibrium properties have been studied rather carefully in the
usual, {\it grand--canonical} Ising model \cite{coni,stauffer}, this topic
seems to have been largely overlooked. Only recently, the
corresponding thermal behaviour of an adatom or vacancy island
of monoatomic height on a crystal surface has been
investigated \cite{selke}. 

In particular, we consider a square lattice
with $L^2$ sites and full periodic boundary conditions. Neighbouring
spins, $S_{i(j)}= \pm 1$, may interact ferromagnetically, with
the coupling term $-J S_i S_j$, $J > 0$. We assume that $N^2$, out of
totally $L^2$, spins are '$-$' spins, with the magnetization
\begin{equation}
M = 1 -2 (N^2/L^2)
\end{equation}
being conserved when varying the temperature, $T$.

The resulting phase diagram in the $(M,T)$--plane is known
to display a transition of first order between
droplet and stripe phases \cite{schlos,zia,neu}. The
transition takes place in the coexistence
region, which is bounded by
$T_0(M_0)$ describing the temperature dependence of the spontaneous
magnetization of the standard Ising model. In
the thermodynamic limit, $M_0$
is determined by \cite{mccoy}
\begin{equation}
M_0 = \left[ 1 - \left( \sinh 2 J/(k_BT_0) \right)^{-4} \right]^{1/8}.
\end{equation}

Another intriguing feature of the droplet phase in the coexistence
region, $T < T_0$, will be discussed in the following, the thermal
stability of the droplet of minority
spins, i.e. '$-$' spins
for $M > 0$.

At $M > 1/2$, the $N^2$ minority spins form a single square droplet in
the ground state. When increasing the temperature at fixed 
magnetization, the largest cluster will, of course, shrink, but
it may be, at
sufficiently low temperatures, still extensive, with the number
of cluster spins, thermally
averaged, $<N_c>$, being proportional to $L^2$
($\propto N^2$). Accordingly, the droplet size $n_c$ defined by  
\begin{equation}
 n_c = <N_c>/N^2,
\end{equation}
will be non--zero in the
thermodynamic limit, $L, N \longrightarrow \infty$, with $M$ ($N/L$)
being constant.

\begin{figure}
\centerline{\psfig{figure=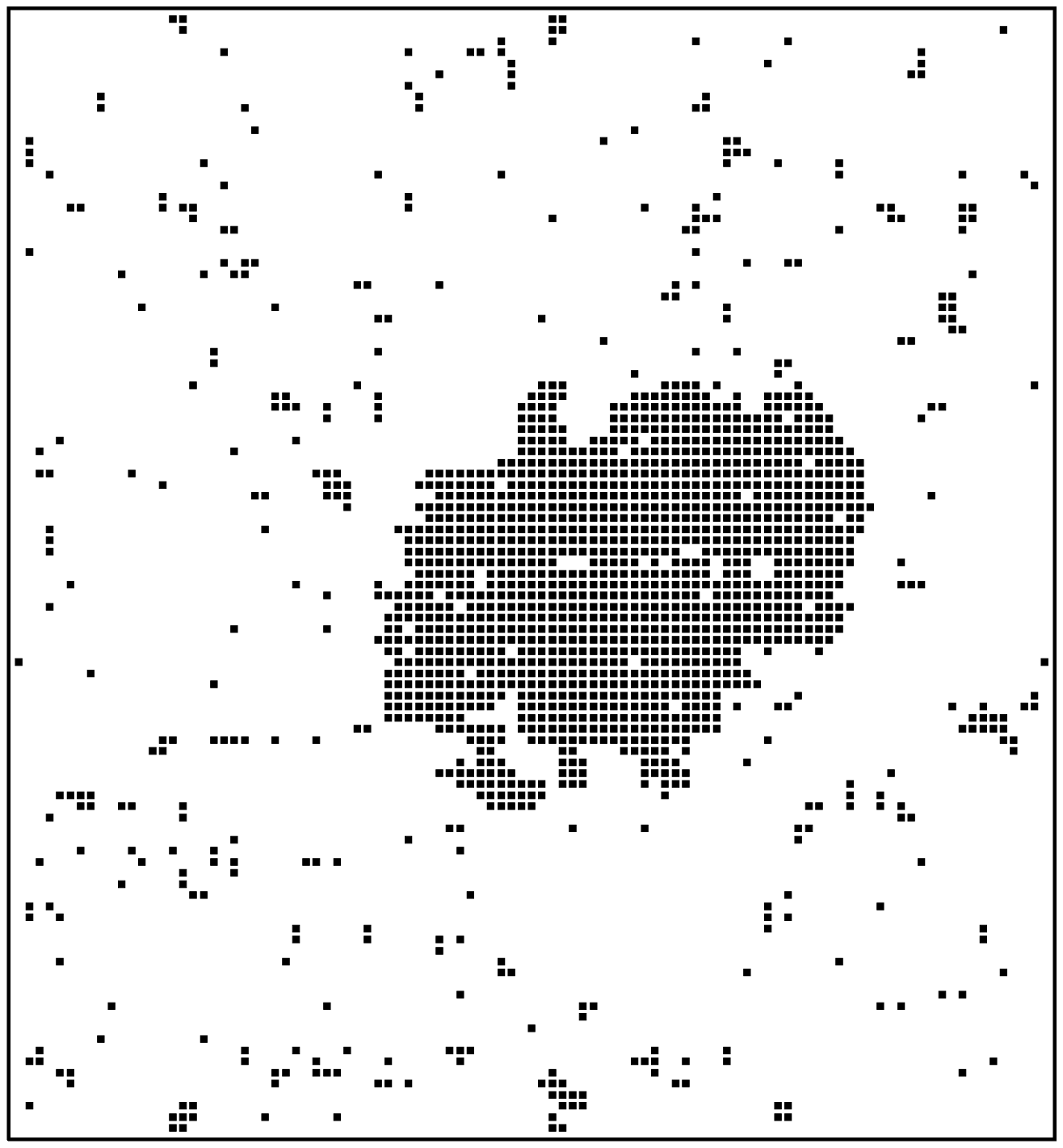,width=8.0cm,angle=270}}
\vspace*{1cm}
\centerline{\psfig{figure=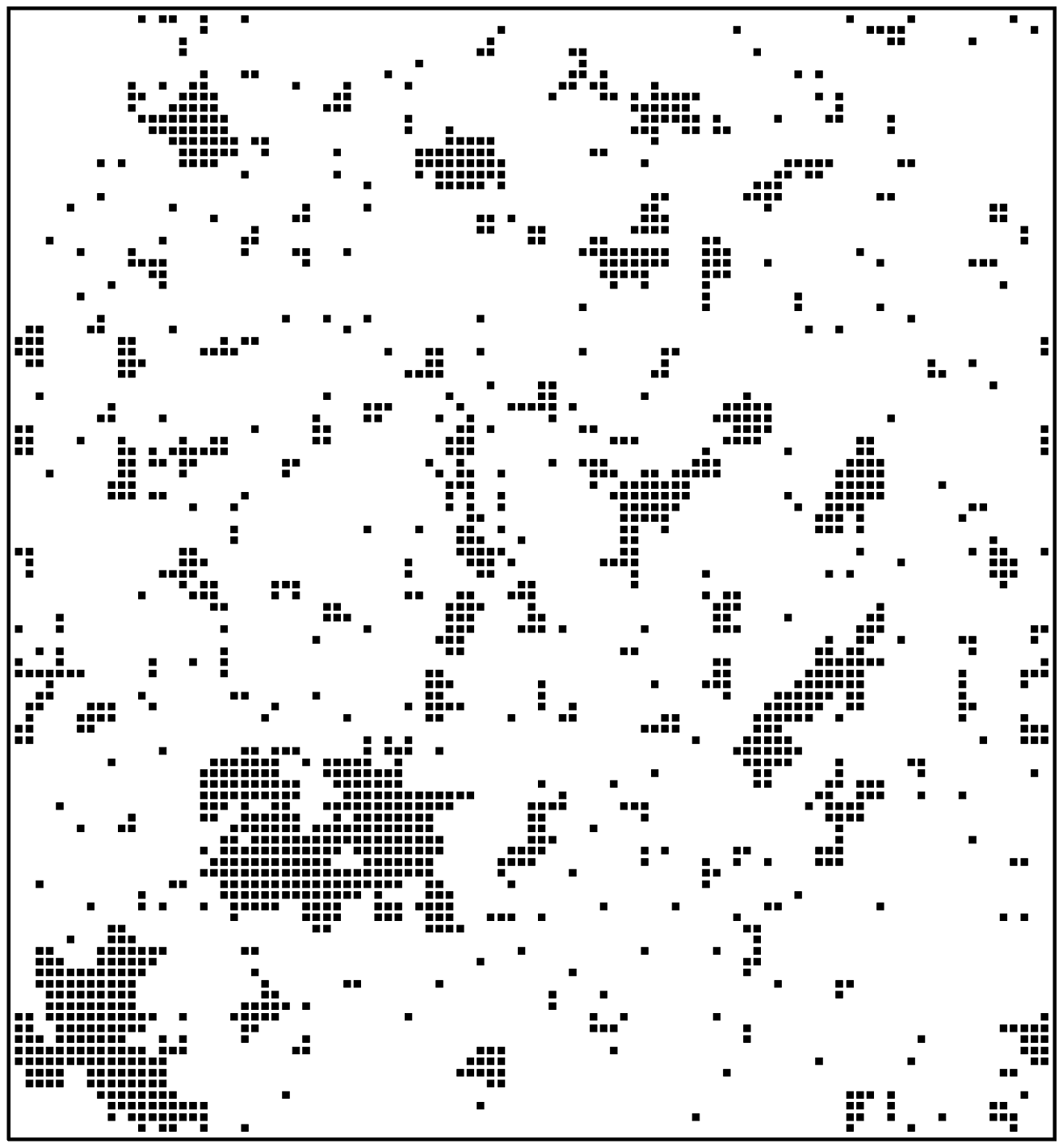,width=8.0cm,angle=270}}
\caption{Typical equilibrium configurations of the two--dimensional
Ising model, at $M=0.68$, below (a: $k_BT/J= 2$) and above
(b: $k_BT/J= 2.25$) the cluster transition, simulating a lattice
with $100^2$ sites; '$-$' spins are denoted by dark squares.}
\end{figure}

However, at and above the coexistence line, $T > T_0$, the largest
cluster is expected to be non--extensive \cite{coni}, see
Fig. 1. Thence, one
expects a 'cluster transition', $T_{cl}$ at or below
$T_0$. Indeed, assuming that, in the coexistence region, see Ref. 8,
\begin{equation}
 - M_d =  M_s = M_0
\end{equation}
where $M_d$ is the magnetization in the droplet, $M_s$ is
the magnetization in the rest of the system, and $M_0$ is given
by Equation (2), one may easily obtain $n_c(T)$, at
fixed magnetization $M$, as   
\begin{equation}
 n_c(T) = (1 + M_0)( M_0 - M)/(2 M_0 (1-M))
\end{equation}
Obviously, the droplet size will then vanish linearly in the
reduced temperature $t= |T_0 - T|/T_0$ as
$T \longrightarrow T_0$ \cite{vb}. 

To investigate the cluster transition and to check this relation, we
did Monte Carlo simulations using a nonlocal spin-exchange
algorithm \cite{barkema} as well as an efficient
cluster algorithm \cite{bloete}, studying square lattices of
various sizes, $L^2$, with $L$ ranging from 25 to 950, at various
magnetizations $M$, ranging from 0.68 to 0.98.

As illustrated in Fig. 2, the droplet size $n_c(M,L,T)$ seems
to approach Equation (5), as one
considers larger and larger lattices, being presumably
valid in the thermodynamic limit, $L, N \longrightarrow \infty$. This
behaviour holds at all values of $M$, we studied. In
addition, one observes rather interesting finite--size effects.

\begin{figure}
\centerline{\psfig{figure=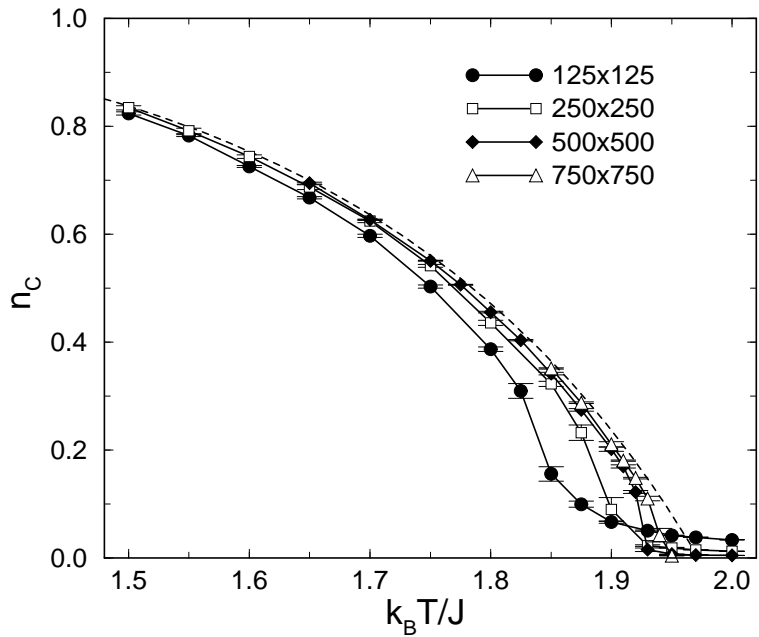,width=10.0cm,angle=270}}
\caption{Droplet size $n_c$ as function of temperature for
lattice sizes $L$ ranging
from 125 to 750, at fixed magnetization $M=0.92$. Error bars stem from 
averaging over at least $N= 10$ Monte Carlo runs
with different random numbers. The dashed
line corresponds to Equation (5).}
%\label{fig2}
\end{figure}

First, the cluster transition is signalled, in a finite
system, by the turning point of $n_c(T)$, at $T_s(L)$. For
moderate system sizes, with $L$ up to, say, 750, $T_s(L)$ is
found to vary (almost) linearly in $1/L$, for different values
of $M$. A straightforward linear extrapolation would
yield cluster transition temperatures close to, but somewhat
below $T_0$. Accordingly, one may expect subtle finite--size corrections
to the observed linear behaviour.-- In this context, attention
is drawn to the non--monotonic size dependence of $n_c(T)$
close to the coexistence temperature $T_0$, see Fig. 2. Obviously, rather
large systems need to be considered in the vicinity of $T_0$ to
allow a reliable extrapolation to the thermodynamic limit.

Another interesting finite--size effect is found by studying 
the size dependence of the largest cluster $<N_c>$ close
to $T_0$. For small lattices, the droplet size scales like
$L^x$, with the (effective \cite{pl}) exponent $x$, at fixed
temperature, being quite
small, e.g. at $M= 0.92$ $x$ is seen to be around 0.5 to 0.7.
Only for larger lattices, the cluster seems to become
more compact, with $x \approx 2$. The crossover, at $L_c$, is
characterised by a pronounced peak in the effective exponent
$x$, with the peak height and crossover length $L_c$ increasing
as $T$ approaches $T_0$. Obviously, this phenomenon is not
predicted by Equation (5) which assumes a compact, extensive
droplet, as seems to be correct in the thermodynamic limit.

Previous evidence for the cluster transition is rather scarce. In the
context of an adatom island on a crystal surface, the
related transition was noticed \cite{selke}, but it was not
discussed in the framework of the phase diagram of the corresponding Ising
model. In that study, a non--trivial critical exponent describing
the vanishing of the droplet size $n_c$ on approach to
the cluster transition had been obtained, estimating the
transition temperature from a straightforward linear
extrapolation of simulational data for moderate
system sizes, as discussed above.-- A possible hint on the
cluster transition might be hidden in 
a renormalization group study of Saito \cite{saito} on quenches
in Ising models with conserved magnetization, finding, however,
a transition below the coexistence line, without
mentioning of the thermal stability of the droplet.-- Attention is also
drawn to the work of Kert\'{e}sz on droplet stability in Ising
models in an external field \cite{ker} as well as related recent
work on Ising cubes \cite{car,gul} and cluster shapes \cite{vel}.

It looks quite promising to analyse various cluster properties close
to the cluster transition, in
analogy to percolation studies \cite{stauffer}. For
instance, we find
that the second moment of the cluster size distribution, excluding the largest
cluster, shows a pronounced peak close to $T_{s}$, indicating
a divergence at the cluster transition in the thermodynamic
limit. Furthermore, the specific
heat also displays a pronounced maximum close to $T_{s}$. 

In summary, we conclude that there seems to be, in
the two--dimensional Ising model with conserved magnetization,
a cluster transition
at the coexistence line, at which
the droplet of minority spins looses, in the thermodynamic
limit, its extensivity. We think that the interesting
cluster and thermal properties as well as finite--size
effects close to that transition deserve to
be studied in more detail in the future, both in two and three
dimensions.

\ack
Useful discussions with H. van Beijeren, J. Hager, J. Schmelzer jr.,
D. Stauffer, and Y. Velenik are
gratefully acknowledged. M. P. thanks the
Deutsche Forschungsgemeinschaft for financial
support.

\end{document}